# Scaling the *h*-index for different scientific ISI fields

## Juan E. Iglesias* and Carlos Pecharromán+

*Instituto de Cerámica y Vidrio, Consejo Superior de Investigaciones Científicas, Cantoblanco, 28049, Spain.
+Instituto de Ciencia de Materiales, Consejo Superior de Investigaciones Científicas, Cantoblanco, 28049, Spain.

*Corresponding Author: e-mail: jeiglesias@icv.csic.es

**Abstract**

We propose a simple way to put in a common scale the *h* values of researchers working in different scientific ISI fields, so that the previsible misuse of this index for inter-areas comparison might be prevented, or at least, alleviated.

**Keywords:**

## 1. Introduction

The proposal by Hirsch (2005a,b) of a single index, *h*, to characterize the significance of the scientific output of a researcher has stirred a wave of comment of planetary proportions, mainly appreciative reactions, such as that of Ball (2005), Bornmann & Daniel (2005) Frangopol (2005), and Imperial & Rodríguez-Navarro (2005). The Hirsch index of a researcher is the highest integer *h* such that *h* among this person's $N_p$ papers have collected at least *h* citations, while the remaining $N_p$-*h* papers have less than *h* citations each. Hirsch convincingly argues that *h* is better than any other single index that one can reasonably come up with, such as: a) total number of papers $N_p$; b) total number of citations, $C_{total}$; c) average value of citations per paper, $C_{total}/N_p$; d) number of "significant papers", i.e. papers with more than a prescribed number *p* of citations; e) number of citations to each of the *q* most cited papers (the last two are really two-parameter indices).

The question why anybody would want to classify people by just a single number is scarcely touched upon, probably because this author considers it obvious that such a course is inescapable; and one can conclude in view of the frenzy that the new index is stirring, that it is just possible that he might be absolutely right. One can speculate that this bare-bones, one-dimensional way of ranking scientists may not be completely unrelated with the predominantly Anglo-Saxon habitude of classifying base-ball and cricket players by their batting average (see, for instance, http://en.wikipedia.org/wiki/Batting_average), basketball performers (http://www.nba.com/statistics/players/scoring.jsp) by their rebounds statistics, and the like. Physicists are particularly fond of models involving a minimum of parameters, and this approach is well anchored in the celebrated principle of Ockham's razor; however, one ought to remember in this connection that Nature has no obligation to us of being simple, a statement that can be proved by the elementary realization that, were it otherwise, Quantum Mechanics would not be around.

Single-index evaluation makes citation analysis come to mind, where it is barely remembered that Garfield originally defined two indices to classify journals, the impact factor and the citation half-life. Darwinian evolution has primed absolutely the impact factor and nobody cares anymore about a journal citation half-life, with the resulting emergence as primary sources of scientific results of journals whose vocation was to carry weekly science news and mild amusement (see, for instance, McManus, 1976) to the educated person, and had correspondingly negligible values for the cited half-life, and for the degree of consolidation of the scientific results thus presented. Many specialized journals catering to limited subsets of workers, whose cited half-life was long, indicating that only final, well analyzed results were to be found in them, have been the losers in this race for the impact factor alone, among them many journal sponsored by the learned societies.

Hirsch is careful to point out the perils of using simple-mindedly one single parameter to rank people, particularly if one uses it in "life-changing decisions, such as the granting or denying of tenure". He warns, for instance, that a scientist whose papers have many co-authors will be treated "overly kindly" by the *h* index, and he suggests that in fields where papers with many authors are the rule (such as high-energy experimental physics) some renormalization may be necessary (Batista et al., 2005). However, this kind of data is not directly available from ISI databases, so one of the appealing features of the *h*-index, its easy extraction of the ISI database, would be lost if one had to spend time figuring out how to correct it. The same can be said concerning self-citations: Hirsch contends that their effect on the *h* index is small, but the fact that relatively high values of *h* are, almost invariably, associated with long tails of low-citation papers seems to point out in the opposite direction.

In spite of all the previous caveats and qualifiers, it should be obvious that the *h*-index is going to be used to hand promotions and to allocate resources, despite (Ball, 2005) some research managers pious declarations that they purposely "avoid using impact factors and citation indices". And since the depths of politicians' (and administrators') "uniquely simple personalities" (Caprara, Barbaranelli & Zimbardo, 1997) only begins to be fathomed, it appears necessary to elaborate a bit on the portability of the *h*-index across the boundary between different scientific fields, lest it be grossly misused by eager specimens of the above sets.

That the *h*-index cannot be used off-hand to compare research workers of different areas has been pointed out by Hirsch himself, by noting that the most highly cited scientists for the period 1983-2002 in the life sciences had *h* values that were almost twice those of the most cited physicists; and from a list of the 36 inductees in the US National Academy of Sciences in the biological and biomedical sciences he extracts the same trend, although perhaps with smaller relative differences with respect to the physical sciences. In this paper we suggest a rational method to account for this, introducing a simple multiplicative corrections to the *h* index which depend basically on the ISI field the worker is in, and to some extent, on the number of papers the researcher has published. We propose below a list of these normalizing factors, so the corrected *h* remains relatively simple to obtain.

## 2. The distribution function of citations

The citation distribution function, i.e. the function *N*(*x*) giving the number of papers which have been cited a total of *x* times has not received much attention in the near past, despite the fact that citation data has been increasingly used for the evaluation of scientific productivity of individuals and institutions. Laherrère & Sornette (1998) studied the citation record of the 1120 most cited physicists over the period 1981-1997, in a long paper in which they set out to prove that many distributions generally regarded to be adequately described by a power law were really better described by a stretched exponential; their paper also deals with radio and light emissions from galaxies, oil reserve sizes, urban agglomeration sizes, currency-exchange price variations, species extinction rates, earthquake distribution and temperatures at he South Pole over the last 400000 years. Instead of searching for the function $N(x)$, they ranked the 1120 physicists according to their total number of citations, and plotted this number of citations, $C$, vs. rank, $k$, in what amounts to a Zipf plot. They got a good fit for the Zipf function $C(k)$ assuming it to be a stretched exponential

$$C(k) \propto \exp\left[-(k/k_0)^\beta\right] \qquad (1)$$

with $\beta \cong 0.3$. Redner (1998) attacked this problem by studying several data bases, prominent among them the citation distribution of 783,339 papers published in 1981, and the corresponding 6,716,198 citations to these papers between 1981 and mid-1997 (Small & Pendlebury, 1997). Interestingly, 47% of these papers were uncited, 76% of them had 7 or fewer citations, and only 1% had more than 105 citations. Redner notes that while the curvature which can be appreciated for this data in a log-log plot would indicate a distribution for $N(x)$ of the type of (1), this function is not appropriate to adequately fit the large-$x$ data. Instead, he finds that a good fit can be obtained for the Zipf function $C(k)$ with a power law,

$$C(k) = k_0 k^{-\alpha} \qquad (2)$$

with an exponent $\alpha \approx 1/2$, and for the distribution $N(x)$, with the same functional form with exponent $\alpha \approx 3$. Redner is careful to warn that in the very large $x$ (or very low rank $k$) the data is quite intractable because it is dominated by fluctuations.

## 3. Calculation of the *h* index

### 3.1. Power law distribution

We assume, by analogy with Redner results, that the citation distribution for an average research worker in a given field is given by the Zipf function (2), with a negative exponent. Then,

$$\int_1^{N_p} C(k)\,\mathrm{d}k = \int_1^{N_p} k_0 k^{-\alpha}\,\mathrm{d}k = C_{total} \qquad (3)$$

where $C_{total}$ is the total number of citations and $N_p$ the total number of papers, of that research worker. From this,

$$k_0 = (1-\alpha)\chi N_p^\alpha \qquad (4)$$

where we assume that the number of papers is large compared with unity, and where

$$\chi = C_{total} N_p^{-1} \qquad (5)$$

is the averaged number of citations per paper for that emblematic research worker. The $h$ index of that typical worker is given by the abscissa of the intersection of the Zipf distribution (2) with the line $C(k) = k$. Hence,

$$h = \left[(1-\alpha)\chi N_p^\alpha\right]^{\frac{1}{(\alpha+1)}} \qquad (6)$$

We follow Redner (1998) and use $\alpha = 1/2$ to get

$$h = \sqrt[3]{\frac{N_p}{4}} \chi^{2/3} \qquad (7)$$

It appears reasonable to assume for the average worker in a given scientific field the world average of citations/paper which corresponds to that particular field, and we elaborate on this below. But notice also that the $h$ value depends on the cubic root of the total number of papers as well. Hence Eq (7) permits easy comparison of people whose total number of papers differs by no more than about 50%.

### 3.2. Stretched exponential distribution

The stretched exponential function was originally used by Hirsh as a model of Zipf function. We show below that in this case there is a relationship between the total number of papers, the average number of citations per paper in the area of work, and the $h$ factor. The analytic variation of $h$ is more intricate here than it is in the case of a power-law function. We assume

$$C(k) = k_0 e^{-\eta k^\beta} \qquad (8)$$

where $k_0$ and $\eta$ are coefficients to be determined, while $\beta$ has been found to be approximately $\beta = 0.3$ (Laherrère & Sornette, 1998; Redner, 1998).

The total number of citations, $C_{total}$ is given by:

$$C_{total} = \sum_1^{N_p} C(k) \simeq \int_0^\infty C(k)\,\mathrm{d}k = \int_0^\infty k_0 e^{-\eta k^\beta}\,\mathrm{d}k \qquad (9)$$

In this case we suppose that the total number of papers is large enough for us to extend the upper limit of the integral to infinity, at the cost of slightly overestimating the total number of citations. Under this approximation, the integral can be written analytically as:

$$\int_0^\infty e^{-\eta k^\beta}\,\mathrm{d}k = \frac{\Gamma\left(1+\frac{1}{\beta}\right)}{\eta^{1/\beta}} \qquad (10)$$

where $\Gamma(z)$ is the usual gamma function (see Spanier & Oldham, 1987).

Hence, the distribution function can be written as:

$$C(k) = C_{total} \frac{\eta^{1/\beta}}{\Gamma\left(1+\frac{1}{\beta}\right)} e^{-\eta k^\beta} \qquad (11)$$

The distribution function given by equation (8) gives zero citations only when the rank is infinity. We assume that the least cited paper has rank $fN_p$ so that $C(fN_p) = 1$, from which:

$$1 = \chi_c N_p \frac{\eta^{1/\beta}}{\Gamma\left(1+\frac{1}{\beta}\right)} e^{-\eta f^\beta N_p^\beta} \qquad (12)$$

where $f$ is the fraction of papers which have been cited at least once, and $\chi_c$ is the average number of citations for those papers.

This expression can be treated as a transcendental equation in $\eta$. In fact:

$$\beta \left[ \frac{f \, \Gamma\left(1+\tfrac{1}{\beta}\right)}{\chi_c} \right]^{\beta} = \left(\beta \eta f^{\beta} N_p^{\beta}\right) e^{-\left(\beta \eta f^{\beta} N_p^{\beta}\right)} = x e^{-x} \quad (13)$$

The solutions of this equation for $x = \beta\,\eta(fN_p)^{\beta}$ can be obtained numerically. However, because the maximum of the function $xe^{-x}$ is $e^{-1}$, equation (13) only has solutions if

$$\chi_c > f \, \Gamma\left(1+\tfrac{1}{\beta}\right)(e\,\beta)^{\tfrac{1}{\beta}} \quad (14)$$

In the case of $f=1$ and $\beta=0.3$, it results that $\chi_c>4.69$. However, we have found that computation with $\chi_c$ values smaller than 8 may produce considerable numerical errors in the calculation of the $\eta$ parameter.

Once $\eta$ is known (which only depends on $C_{total}$ and $\beta$), the value of $h$ can be determined. The definition of the Hirsch index is given by:

$$h = C(h) \quad (15)$$

and then, using equation (11) and the result of equation (13) we get:

$$h = C_{total} \frac{\eta^{1/\beta}}{\Gamma\left(1+\tfrac{1}{\beta}\right)} e^{-\eta h^{\beta}} \quad (16)$$

Using the same procedure as in equation (13), equation (16) can be transformed into:

$$\frac{\Gamma\left(1+\tfrac{1}{\beta}\right)^{\beta}}{\beta C_{total}^{\beta} \eta^2} = \frac{e^{-\beta \eta h^{\beta}}}{\beta \eta h^{\beta}} = \frac{e^{-z}}{z} \quad (17)$$

so that, following a procedure as in Eq (13), a numerical solution is obtained for $z = \beta\,\eta\,h^{\beta}$.

Although it can be seen that the $h$ index depends both on the number of published papers $N_p$ and on the average number of citations per paper, $\chi_c$, in the case of the stretched exponential distribution the explicit dependence of $h$ on $N_p$ and $\chi_c$ is not easy to determine for the whole range of values, and we cannot assume a slow variation respect to the number of papers, $N_p$.

### 4. Normalization data

We use the ISI average number of citations/paper for each scientific field
(http://portal.isiknowledge.com/portal.cgi?DestApp=ESI&Func=Frame,
contained in Essential Science Indicators, Baselines. The data (downloaded Jan, 2006) is reproduced in Table I.

**Table I**
Average number of citations/paper as of Dec, 2005, in the different ISI fields, of papers published in each year
(data from ISI, http://portal.isiknowledge.com/portal.cgi?DestApp=ESI&Func=Frame; downloaded Jan, 2006)

| Fields | 1995 | 1996 | 1997 | 1998 | 1999 | 2000 | 2001 | 2002 | 2003 | 2004 | 2005 | All Years |
|---|---|---|---|---|---|---|---|---|---|---|---|---|
| Agricultural Sciences | 8.36 | 8.13 | 7.61 | 7.38 | 6.78 | 6.13 | 4.85 | 3.53 | 2.29 | 0.91 | 0.15 | **4.93** |
| Biology&Biochemistry | 26.52 | 24.5 | 24.45 | 21.81 | 19.5 | 17.37 | 14.14 | 10.54 | 6.77 | 3.04 | 0.48 | **15.37** |
| Chemistry | 13.57 | 12.94 | 12.31 | 11.82 | 10.67 | 9.78 | 7.84 | 6.34 | 4.11 | 1.95 | 0.33 | **8.09** |
| Clinical Medicine | 19.13 | 17.27 | 16.31 | 15.08 | 13.7 | 12.03 | 9.96 | 7.66 | 4.94 | 2.16 | 0.37 | **10.58** |
| Computer Science | 5.03 | 4.93 | 4.8 | 4.7 | 4.05 | 3.31 | 3.08 | 2.6 | 1.19 | 0.47 | 0.09 | **2.49** |
| Economics & Business | 9.23 | 7.5 | 7.22 | 6.12 | 5.11 | 4.2 | 3.16 | 2.4 | 1.3 | 0.52 | 0.11 | **4.17** |
| Engineering | 5.43 | 5.14 | 5.21 | 4.58 | 4.23 | 3.71 | 3.14 | 2.27 | 1.43 | 0.63 | 0.1 | **3.17** |
| Environment/Ecology | 14.63 | 13.9 | 13.15 | 12.44 | 10.77 | 9.56 | 7.23 | 5.3 | 3.29 | 1.36 | 0.22 | **7.81** |
| Geosciences | 15.06 | 14.1 | 13.15 | 12.2 | 10.3 | 8.52 | 6.9 | 4.69 | 3.02 | 1.3 | 0.28 | **7.65** |
| Immunology | 34.12 | 30.67 | 28.76 | 28.24 | 24.18 | 22.15 | 18.53 | 13.71 | 8.85 | 4.16 | 0.57 | **19.55** |
| Materials Science | 7.64 | 7.31 | 6.77 | 6.59 | 5.96 | 5.5 | 4.55 | 3.4 | 2.24 | 0.96 | 0.14 | **4.32** |
| Mathematics | 5.16 | 4.87 | 4.37 | 3.87 | 3.58 | 2.91 | 2.26 | 1.74 | 1.03 | 0.47 | 0.08 | **2.66** |
| Microbiology | 24.33 | 22.65 | 22 | 20.74 | 18.32 | 15.85 | 13 | 9.8 | 6.31 | 2.94 | 0.47 | **14.02** |
| Molecular Biology&Genetics | 42.72 | 39.75 | 38.33 | 35.94 | 32.45 | 28.08 | 23.26 | 17.53 | 11.16 | 5.21 | 0.8 | **24.57** |
| Neuroscience&Behavior | 29.99 | 27 | 25.69 | 23.81 | 21.57 | 18.75 | 15.56 | 11.3 | 6.81 | 2.93 | 0.42 | **16.41** |
| Pharmacology&Toxicology | 16.11 | 14.54 | 14.43 | 12.98 | 12.41 | 11.11 | 9.42 | 7.44 | 4.55 | 2.05 | 0.28 | **9.4** |
| Physics | 12.3 | 11.79 | 10.81 | 10.16 | 9.4 | 8.58 | 7.1 | 5.41 | 3.66 | 1.89 | 0.36 | **7.22** |
| Plant & Animal Science | 11.25 | 10.58 | 9.8 | 8.79 | 7.92 | 6.87 | 5.58 | 4.07 | 2.57 | 1.16 | 0.21 | **6.15** |
| Psychiatry/Psychology | 15.21 | 13.78 | 13.39 | 11.9 | 10.99 | 8.97 | 7.33 | 5.04 | 3.11 | 1.3 | 0.25 | **8.24** |
| Social Sciences, general | 5.97 | 5.72 | 5.48 | 5.12 | 4.53 | 3.91 | 3.03 | 2.29 | 1.35 | 0.6 | 0.17 | **3.46** |
| Space Science | 18.71 | 17.64 | 17.88 | 16.06 | 16.88 | 12.29 | 12.26 | 8.43 | 6.67 | 3.21 | 0.64 | **11.58** |

The data reflects the average number of citations (as of Dec. 2005) a paper in each field and year of publication has received since publication. The value is naturally larger for older papers, but a tendency to stabilize is visible in all fields after a few years. The average values for each year are strongly dependent on the field, and can vary by a factor as high as 9; thus the quotient between the values corresponding to "Molecular Biology & Genetics" and those corresponding to "Mathematics" is 8.3 for papers published in 1995 (we do not consider here the field labelled "Multidisciplinary" by ISI).

**4.1. Power law distribution**

In Figs 1 and 2 we have plotted for each field and year the normalizing factor

$$f_i = \left[ \chi_{Physics} / \chi_i \right]^{2/3} \qquad (18)$$

with $\chi_i$ as defined in Eq. (5). The values for "Physics" are taken as reference to better compare with data in Hirsch. The normalizing factor $f_i$ is the value by which the $h$ index in field $i$ has to be multiplied in order to put it in the same scale as that of the field "Physics", as described in Eq. (7). In Fig 1 we can see that after about six years, the normalizing factor $f_i$ for each field becomes quite stable; a notable exception is the field "Economics & Business", which shows a quite marked descending pattern over the six years shown, and to a less extent, "Mathematics" and "Computer Sciences", which show a descent with some degree of stabilization just in the 1995-1998 interval. A similar pattern can be observed in Fig 2, where the fields with $f_i < 1$ have been plotted. Most fields show considerable stability over the six-year period shown, with the exception of "Geosciences" "Psychiatry/Psychology" and "Space Sciences", which show a descending pattern coming to a region of stabilization in the period 1995-1998.

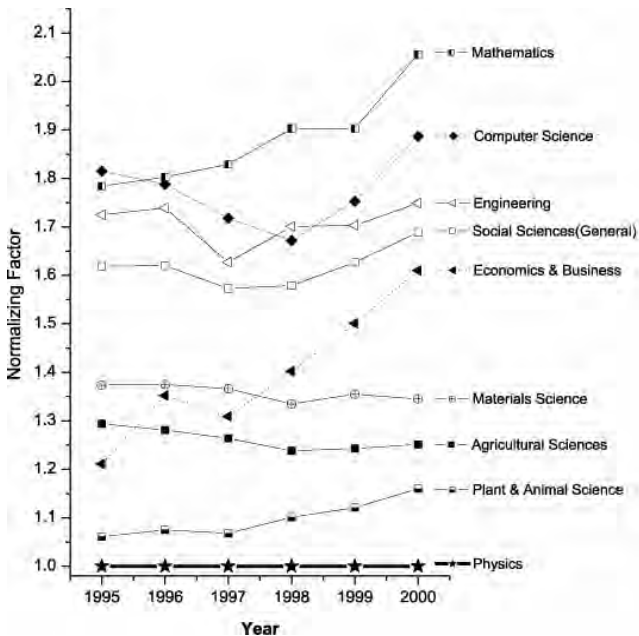

Fig 1.- The values, for each field and year, of the ratio of the values in Table I to those of the field "Physics", raised to the power 2/3, for values of this ratio greater than 1.

The relative variation of each $f_i$ and the range of values it adopts indicate that researchers in the different fields follow different citing patterns. It is remarkable that those fields closely associated with the concept of "experimental science" follow a citing pattern similar to that used by the physicists, although with different citing density scales. Those fields which could be loosely classified as "observational" ("Space Sciences", "Geosciences", "Psychiatry/Psychology", and to some extent, "Social Sciences") seem to follow a qualitatively different citing pattern, with the factor $f_i$ decreasing as the age of the paper increases (between 2000 and 1998). The pattern of "Mathematics" and "Computer Sciences" is more similar to that of the "observational" group than it is to the pattern of the "experimental" ones. Mathematics is not easy to classify, at any rate: some mathematicians (called "formalists") believe that Mathematics are "created", or "invented" in the same sense that Art is, and hence do not have any existence outside the human mind, while for other mathematicians, called "platonists", mathematical truths are external to the human mind, and are out there to be "discovered", or "uncovered" (see, for instance, Livio, 2003). The citation pattern of the field "Economics & Business" (which, in principle, is an "observational" science or a subset of Sociology) is not well behaved for the purposes of this paper, and is the only one for which is difficult to postulate a well-defined normalizing factor.

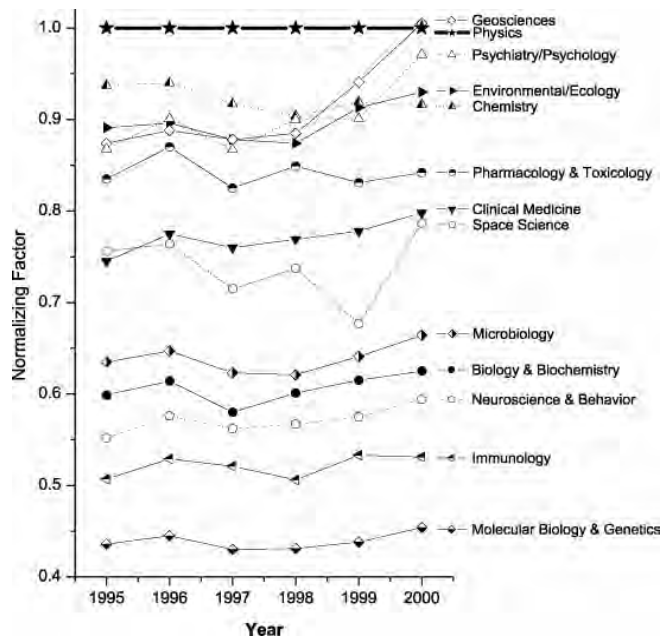

**Fig 2.-** The values for each field and year of the ratio of the values in Table I to those of the field "Physics", raised to the power 2/3, for values of this ratio less than 1.

**4.2. Stretched exponential distribution**

We use the data in Table I as explained below.

**5. Corrections factors**

**5.1. Power law distribution**

We have tabulated in the first column of Table II the recommended values for normalization of the $h$ index, computed under the assumption that the distribution function of the citations is given by Eq (2). They have been calculated by averaging the ratio of the number of citations/paper for each field, for years 1995, 1996, 1997 and 1998, and normalizing to the corresponding values for the field "Physics", prior to applying Eq. (18). Notice that comparison of two researchers having a very different number of papers would require to apply also a correction along the lines of Eq. (7).

It is worth noting that the normalizing factors for the five fields which could be labeled "(molecular) life sciences": "Microbiology", "Biology & Biochemistry", "Neuroscience & Behavior", "Immunology" and "Molecular Biology & Genetics" (see Table II, and Fig 3) agree well with Hirsch's observation that $h$ indices in the life sciences for Nobel prize researchers are about twice those in Physics.

**5.2. Stretched exponential distribution**

The expression of the $h$-index in this distribution model is strongly dependent on the number of published papers of a given author, and so is, consequently, the correction factor. The last four columns of Table II contain the correction factors relative to the field "Physics" for all fields, meant to compare $h$ indices of authors having 100, 200, 500 and 1000 papers. It can be seen that this model predicts correction factors which are more conservative than those of the power law model. It appears that the values predicted by the stretched exponential model converge to those of the power law model in the limit of an infinite number of papers. Some caution must be exercised with ISI fields with citations/paper rates smaller than 8. (Mathematics, Computer Sciences, Engeniering, Social Sciences and Material Sciences), because the numerical method used to calculate the corrector factor may produce some inaccuracies.

**Table II**
**Normalization factor for the ISI Fields of Science, relative to the field "Physics"**

To put $h$-indices of different fields in a common scale, multiply by $f_i$, the tabulated value. The first column gives $f_i$ values calculated from a power-law Zipf plot (for comparison of authors having different number of papers, see text). The remaining columns give correction factors computed under the assumption that the citation distribution function is a stretched exponential, for comparison of authors having a similar number of published papers.

| ISI Fields | Power Law | Stretched Exponential 100 papers | 200 papers | 500 papers | 1000 papers |
|---|---|---|---|---|---|
| Agricultural Sciences | 1.27 | 1.20 | 1.24 | 1.30 | 1.35 |
| Biology & Biochemistry | 0.60 | 0.77 | 0.73 | 0.68 | 0.64 |
| Chemistry | 0.92 | 0.95 | 0.94 | 0.93 | 0.92 |
| Clinical Medicine | 0.76 | 0.86 | 0.83 | 0.80 | 0.77 |
| Computer Science | 1.75 | 1.97 | — | — | — |
| Economics & Business | 1.32 | 1.23 | 1.28 | 1.36 | 1.42 |
| Engineering | 1.70 | 1.79 | — | — | — |
| Environment/Ecology | 0.88 | 0.93 | 0.92 | 0.90 | 0.88 |
| Geosciences | 0.88 | 0.93 | 0.91 | 0.89 | 0.88 |
| Immunology | 0.52 | 0.73 | 0.68 | 0.63 | 0.58 |
| Materials Science | 1.36 | 1.29 | 1.35 | 1.44 | — |
| Mathematics | 1.83 | — | — | — | — |
| Microbiology | 0.63 | 0.79 | 0.75 | 0.71 | 0.67 |
| Molecular Biology&Genetics | 0.44 | 0.68 | 0.64 | 0.57 | 0.53 |
| Neuroscience&Behavior | 0.56 | 0.75 | 0.71 | 0.66 | 0.62 |
| Pharmacology&Toxicology | 0.84 | 0.90 | 0.89 | 0.86 | 0.85 |
| Physics | 1.00 | 1.00 | 1.00 | 1.00 | 1.00 |
| Plant & Animal Science | 1.08 | 1.05 | 1.06 | 1.07 | 1.08 |
| Psychiatry/Psychology | 0.88 | 0.93 | 0.91 | 0.90 | 0.88 |
| Social Sciences, general | 1.60 | 1.58 | 1.72 | — | — |
| Space Science | 0.74 | 0.85 | 0.82 | 0.79 | 0.76 |

**6. Examples taken from Spanish research workers**

**6.1. Power law model**

We have looked up those research workers of Spanish institutions listed by Thomson ISI as "Highly cited scientists" at URL http://portal.isiknowledge.com/portal.cgi?DestApp=HCR&Func=Frame, (downloaded Feb 21, 2006), with the condition that they had not very common last names; excluding commonness of last name should introduce no bias in our sample (Hirsch, 2005a,b), and facilitates unambiguous identification of each worker. The results are summarized in Table III, where we have tabulated the values of $h$, $N_p$ and our corrected (assuming a power law distribution) index, $H$:

$$H = h \cdot f_i \qquad (19)$$

with the normalizing factor $f_i$ taken from the first column of Table II.

**Table III**
**Highly cited Spanish scientists**

| ISI Field | Name | h | H | Nc | Np |
|---|---|---|---|---|---|
| Chemistry | A. Corma | 60 | 55 | 12210 | 625 |
| Clinical Medicine | J. Rodés | 84 | 65 | 15644 | 1047 |
| Environment/Ecology | C. M. Herrera | 35 | 32 | 2213 | 106 |
| Immunology | F. Sánchez Madrid | 56 | 29 | 8112 | 235 |
| Mathematics | D. Nualart | 15 | 28 | 892 | 125 |
| " | J. M. Sanz Serna | 21 | 40 | 1282 | 75 |
| " | J. L. Vázquez | 22 | 42 | 1015 | 111 |
| " | E. Zuazúa | 19 | 36 | 821 | 141 |
| Molecular Biology & Genetics | M. Barbacid | 79 | 35 | 17816 | 217 |
| Neuroscience & Behavior | J. M. Palacios | 72 | 41 | 14231 | 540 |
| Physics | M. Aguilar Benítez | 38 | 38 | 7782 | 214 |
| Plant & Animal Science | C. M. Duarte | 38 | 42 | 2944 | 252 |

$N_c$ = number if citations; $N_p$= number of papers

The initial range 15-84 in the $h$ values of Spain top-cited scientists becomes reasonably more homogeneous after correction, 28-65. The criteria ISI employs to include a scientist in the category of "Highly cited scientist" are not known in any detail, but are certainly dependent on the scientific field. This is particularly clear when one looks at the field "Mathematics", in which four Spaniards are included as "highly cited", all having a citation level considerably lower than that present in scientists of other fields. Still it is puzzling to observe in this connection that A. Wiles is not included in this honor roll, despite the fact that he has become, arguably, one the most celebrated mathematicians of the 20th century, after his renowned proof of so-called Fermat's last theorem was known (Wiles, 1995)[1]. In fact, a simple Thomson ISI search assigns him a mere $h$=12 ($H$=22) on a list of just 13 papers. Since it is by no means usual in other fields to publish single-author papers 109 pages long containing the work of several years, it is easy to see why "Mathematics" is a field that has quite specific rules, and probably requires individualized treatment.

---
[1] Our mention of this paper here does not appear to have a chance of improving the $h$-value of this author.

The results in Table III also indicate that it is unlikely that a research worker will have a relatively high $h$ (or $H$) value without having simultaneously a large number of papers. This could be an indication that self-citations play a role in the value of $h$ more important than that acknowledged by Hirsch. However, Hirsch's argument that self-citations to papers having less than $h$ citations are of no effect on the final value of the $h$ index is undeniable, and hence some quantitative research would be necessary to clarify the influence of the number of self-citations on the $h$ index.

**6.2. Stretched exponential model**

In order to illustrate this model we have used the extant information relative to number of published papers and $h$-factor for some of the scientists of ICMM-CSIC. The choice has been based on our ability to disambiguate data for not so uncommon names on the ISI web. Results have been plotted in figure 3 as large dots. In this figure, three curves of $h$ vs. $N_p$ have been plotted, for $\chi$ =12.7, 11.3 and 7.9, i.e. those values corresponding to Chemistry, Physics and Materials Sciences, for the period from 1995 to 1998. It can be seen that many of the sampled scientists cluster around the curves for Physics and Chemistry, but there are several authors whose representative points lie well above the curve, probably setting them apart from the majority of workers in the Institute. One can speculate that the more the actual $h$-value of a given scientist deviates from the statistical estimate of his/her group, the higher the merit or excellence of that scientist. But of course, a major problem that we have ignored so far is that of classifying a scientist's production under one or other of the ISI headings. In the ICMM there are workers whose papers could be classified as "Chemistry", "Physics", "Material Science", and mixtures thereof. There is no big difference between the statistics for "Physics" and that of "Chemistry", but it is clear, after the data in Table II, that a worker whose bulk of scientific publications belongs properly in the field "Materials Science", would need to have his/her $h$-index multiplied by some factor between 1.29 and 1.36 before a meaningful comparison can be made with the other two groups.

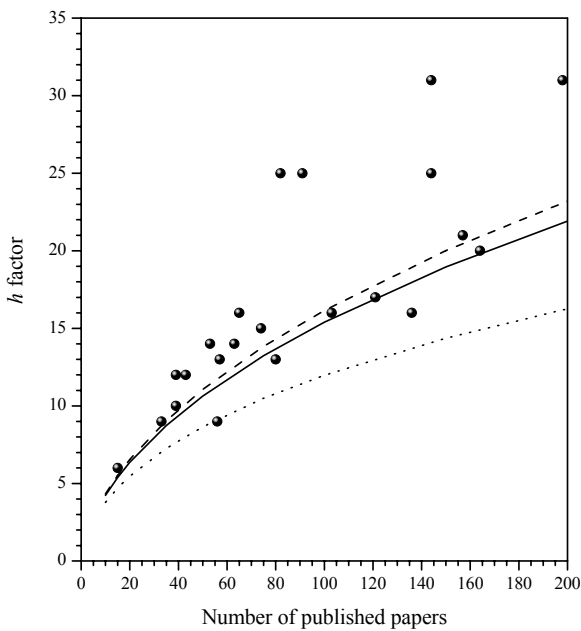

**Fig. 3.-** The $h$ factor vs. number of published paper for several researcher of the ICMM-CSIC. Continuous, dashed and dotted lines represent the "average physics, chemistry and material science standard researcher" according to the stretched exponential model for the Zipf-plot of citations.

We believe this adjustment to the $h$-value is more meaningful than that advanced by Hirsh as the $m$ parameter, i.e., our plot detects, for different numbers of published papers, which scientists clearly deviate from world standards. Thus, a scientist whose representative point is close to the world curve for the adequate field may be thought of as having a worldwide degree of acceptance by his/her peers, independently of the number of published papers being high or low. This provides a fair measure for those people who devote only a part of their time to research, and permits one to compare people who differ in age. According to the present model, after several years (from 5 to 10) from the first publication, the main relevant factors which modify $h$ are the number of citations/paper in the scientific area in which the scientist publishes (which, presumably, only reflects citation habitudes in that particular collective) and the number of published papers.

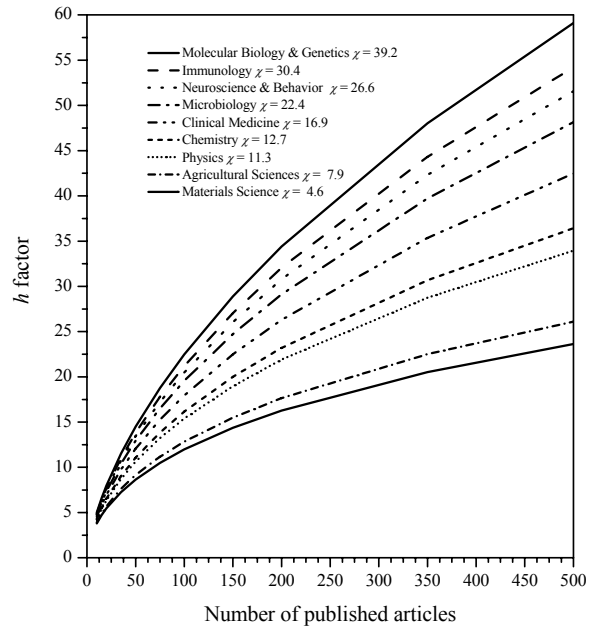

**Fig 4.-** The $h$ factor vs. number of published paper for selected areas of knowledge. Plots have been calculated assuming a stretched exponential model for the Zipf-plot of citations. The following parameters were employed: $f$=0.5, $\alpha$=0.3 and x values were taken from Table I.

**7. Concluding remarks**

We believe the above discussion and data indicate that:

a) Simple-minded comparison of the $h$-index of two people will give meaningless results unless the indices are properly corrected for the fact that different science fields have different citation habitudes, as reflected in the widely differing average values of citations/paper for the different scientific areas. In particular, it is quite obvious that an average scientist should have a number of citations/paper similar to the world average in his/her field, since this will be an indication that this person's papers meet the generally accepted standards of the trade. It could perhaps be argued that a few highly cited scientists could disproportionately inflate that statistic, but then one should remember the sobering fact that about 47% of all published papers go uncited, which makes this distortion extremely improbable. In other words, scientists with a $\chi$ value (cf. Eq (5)) significantly lower than the world average of their field are, in quite a literal sense, substandard, and the same can be said of entire groups showing this kind of individual behavior.

b) In many cases the *h*-index must be corrected for the number of published papers, an effect shown with different intensity by both models used in the calculation; it can be said that, theoretically, any model will show that effect, which is inherent in the Zipf plot. It could be objected that both, number of papers and *h* value are independent indicators of a certain degree of accomplishment, so it should not be surprising that these numbers are, in a way, correlated. But then a high number of papers which is not accompanied by a correlative value of the *h* value would obviously indicate a low quality of the scientific content of these articles; and, on the contrary, a value of the *h*-index higher than that which could be expected from the number of papers, can be taken as an indication of quality when the number of papers is not high, for instance, due to partial dedication to scientific duty or young age.

c) The model predicting a stronger influence of the number of published papers in the *h*-index implicitly predicts an important influence on that index of the number of self-citations, since a high value of this number is unavoidably linked with a high number of papers. Hence, should the future evidence favor Hirsch's opinion that the effect of self-citation on the *h*-index is negligible, that same evidence would appear to favor the power-law model over the stretched exponential model of distribution, since in the former one the number of published papers is only weakly linked to the *h*-number; otherwise, one would have to conclude that the citations (or lack thereof) to papers of a scientist not contributing to the *h*-index could have an influence on those papers that do contribute to that index.